\documentclass[12pt]{article}

\usepackage{scicite}

\usepackage{times}
\usepackage{graphicx}
\usepackage{xspace}
\usepackage[version=4]{mhchem}
\usepackage[colorlinks,
linkcolor=blue,
anchorcolor=blue, 
citecolor=blue,
urlcolor=blue,
]{hyperref}

\newenvironment{sciabstract}{%
\begin{quote} \bf}
{\end{quote}}

\newcounter{lastnote}

\usepackage[separate-uncertainty = true,multi-part-units=single]{siunitx}
\newcommand{\tplus}{\ensuremath{^{3+}}\xspace}

\newcommand{\absVR}{$\lvert V_{\mathrm{R}} \rvert$\xspace}
\newcommand{\deltaabsVR}{$\Delta \lvert V_{\mathrm{R}} \rvert$\xspace}
\newcommand{\erg}{\ensuremath{^4}I\ensuremath{_{15/2}}\xspace}
\newcommand{\ere}{\ensuremath{^4}I\ensuremath{_{13/2}}\xspace}

\title{Photoionization detection of a single Er\tplus ion with sub-100-ns time resolution}

\author
{Yangbo Zhang,$^{1,3}$ Wenda Fan,$^{1,3}$ Jiliang Yang,$^{1,3}$ Hao Guan,$^{1,2,3}$ Qi Zhang,$^{1,3}$\\
Xi Qin,$^{1,2,3}$ Changkui Duan,$^{1,3}$ Gabriele G. de Boo,$^{4}$ Brett C. Johnson,$^{5,6}$\\
Jeffrey C. McCallum,$^{6}$ Matthew J. Sellars,$^{7}$ Sven Rogge,$^{4}$\\
Chunming Yin,$^{1,2,3\ast}$ Jiangfeng Du$^{1,2,3}$\\
\\
\normalsize{$^{1}$CAS Key Laboratory of Microscale Magnetic Resonance and School of Physical Sciences,}\\
\normalsize{University of Science and Technology of China, Hefei 230026, China,}\\
\normalsize{$^{2}$Hefei National Laboratory, University of Science and Technology of China, Hefei 230088, China,}\\
\normalsize{$^{3}$CAS Center for Excellence in Quantum Information and Quantum Physics,}\\
\normalsize{University of Science and Technology of China, Hefei 230026, China,}\\
\normalsize{$^{4}$Centre of Excellence for Quantum Computation and Communication Technology,}\\
\normalsize{School of Physics, University of New South Wales, NSW 2052, Australia,}\\
\normalsize{$^{5}$Centre of Excellence for Quantum Computation and Communication Technology,}\\
\normalsize{School of Engineering, RMIT University, Victoria 3001, Australia,}\\
\normalsize{$^{6}$Centre of Excellence for Quantum Computation and Communication Technology,}\\
\normalsize{School of Physics, University of Melbourne, Victoria 3010, Australia,}\\
\normalsize{$^{7}$Centre of Excellence for Quantum Computation and Communication Technology,}\\
\normalsize{Research School of Physics and Engineering, Australian National University, ACT 0200, Australia,}\\
\\
\normalsize{$^\ast$To whom correspondence should be addressed; E-mail:  Chunming@ustc.edu.cn.}
}

\date{}

\begin{document} 


\baselineskip24pt


\maketitle 


\begin{sciabstract}
  Efficient detection of single optical centers in solids is essential for quantum information processing, sensing, and single-photon generation applications. In this work, we use radio-frequency (RF) reflectometry to electrically detect the photoionization induced by a single Er\tplus ion in Si. The high bandwidth and sensitivity of the RF reflectometry provide sub-100-ns time resolution for the photoionization detection. With this technique, the optically excited state lifetime of a single Er\tplus ion in a Si nano-transistor is measured for the first time to be \SI{0.49(4)}{\us}. 
  Our results demonstrate an efficient approach for detecting a charge state change induced by Er excitation and relaxation. This approach could be used for fast readout of other single optical centers in solids and is attractive for large-scale integrated optical quantum systems thanks to the multi-channel RF reflectometry demonstrated with frequency multiplexing techniques.
\end{sciabstract}


\section*{Introduction}

Efficient detection is key to the discovery and utilization of single optical centers in solids. Since the first observation on a single-center basis\cite{gruber_scanning_1997}, nitrogen-vacancy (NV) centers in diamond have been used for quantum information processing\cite{togan_quantum_2010,pfaff_unconditional_2014}, single-photon generation\cite{smith_single_2020}, and quantum sensing under ambient conditions\cite{maze_nanoscale_2008,balasubramanian_nanoscale_2008}. The success is built upon strong, spin-dependent fluorescence as well as cyclic optical transitions\cite{robledo_high_fidelity_2011}. Single optical centers possessing similar properties have also been reported in other solids, including common semiconductor materials such as SiC\cite{castelletto_silicon_2014,lohrmann_single_photon_2015,christle_isolated_2015,widmann_coherent_2015,anderson_electrical_2019} and Si\cite{G_center,W_center,T_center}. More recently, detection and coherent control of single rare-earth ions in solids have been achieved by using cavity coupling\cite{Dibos_Atomic_2018,zhong_optically_2018}, which is essential to enhance and efficiently collect the photon emission from single rare-earth ions. In these above experiments, the number of collected photons is used as the readout signal. Therefore, the readout fidelity is limited by spin-flip errors, which occur during repeated excitations for getting sufficient signal contrast, and by the photon collection efficiency, despite the improvement due to the use of a solid-immersion lens\cite{siyushev_monolithic_2010,hadden_strongly_2010}.

Recent implementation of spin-to-charge conversion provides spin readout fidelity over \SI{98}{\percent} for NV centers in diamond\cite{zhang_high_fidelity_2021,irber_robust_2021} and divacancy centers in SiC\cite{anderson_five_second_2022}. This is a promising path towards deterministic readout; however, the final charge state readout has a limited speed as it still relies on photon collection within a time window on the order of 1-\SI{10}{\ms}. In contrast, electrical charge state readout, widely used in nano-electronic devices, offers high fidelity with short integration times\cite{elzerman_single-shot_2004,morello_single-shot_2010}. This technique has been demonstrated for single Er\tplus ions in Si\cite{Yin_Optical_2013,hu_time-resolved_2022,yang_spectral_2022}, but the readout speed in these studies\cite{hu_time-resolved_2022,yang_spectral_2022} was limited by the bandwidth of the current measurement. The readout speed could be significantly increased by adopting radio-frequency (RF) reflectometry\cite{gonzalez-zalba_probing_2015,vigneau_probing_2022}, a fast and sensitive charge detection technique. This technique was first established on Al-based single electron transistors with a bandwidth exceeding \SI{100}{\MHz}\cite{schoelkopf_radio-frequency_1998} and was later applied to semiconductor qubits\cite{crippa_gate-reflectometry_2019,jirovec_singlet-triplet_2021}. It is an attractive readout technique for large-scale integrated quantum systems, as it can be performed on devices with only a single electrical contact\cite{house_high_sensitivity_2016} and also allows multi-channel parallel readout using frequency multiplexing\cite{laird_coherent_2010,house_high_sensitivity_2016}.

Here, we report photoionization detection of a single Er\tplus ion in Si using RF reflectometry. We first introduce the experimental design focusing on the high-bandwidth RF reflectometry technique and then investigate the optically excited state lifetimes of two single Er\tplus ions. Finally, we analyze the sensitivity and limits of RF detection and discuss future optimization and other potential applications of this technique.

\section*{Results}

\begin{figure}
    \centering
    \includegraphics[width=0.5\textwidth]{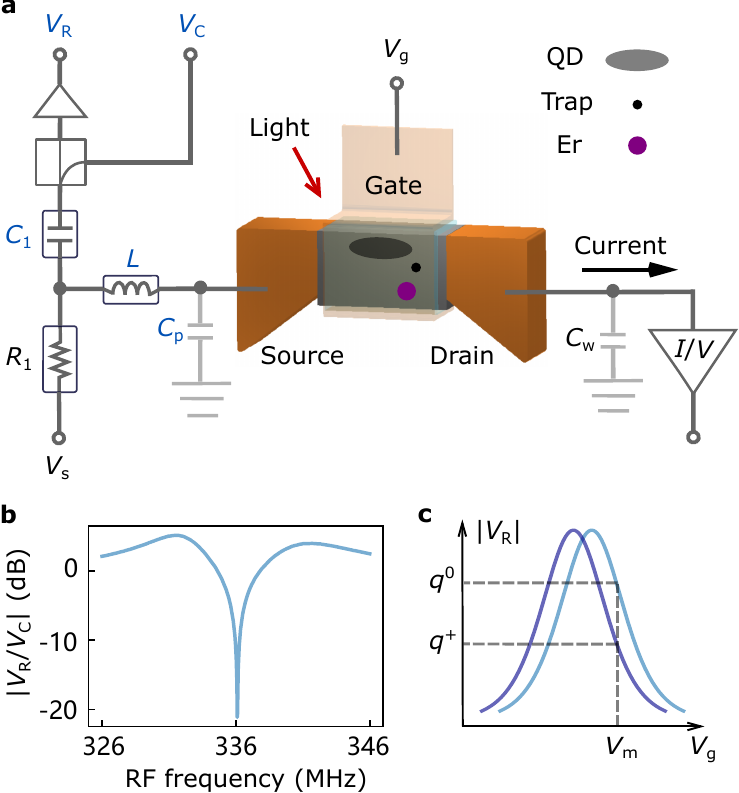}
    \caption{\textbf{Experimental setup.} 
    (\textbf{a}), Circuit diagram showing both RF and DC connections to the device. The DC current is measured from the drain of the device. Its bandwidth is limited by the input impedance of the current pre-amplifier ($I/V$), the device resistance and the wiring capacitance $C_{\mathrm{w}}$. For the RF measurement, a resonant circuit is connected to the source of the device, consisting of a commercial inductor ($L=\SI{470}{\nano\henry}$) and the parasitic capacitance ($C_{\mathrm{p}}= \SI{490.1}{\femto\farad}$). A carrier RF signal ($V_{\mathrm{C}}$) is sent to the resonant circuit, and the reflected signal ($V_{\mathrm{R}}$) from the resonant circuit can be used to detect an impedance change of the device. A bias tee made up of $R_{\mathrm{1}}$ and $C_{\mathrm{1}}$ separates the DC source voltage $V_{\mathrm{s}}$ and the RF signal so that the DC and RF measurements can be carried out simultaneously. In the FinFET channel, the implanted Er\tplus ion, a trap and a QD involved in the detection mechanism are displayed.
    (\textbf{b}), Characteristic of the resonant circuit when all three terminals of the device are grounded. The curve reveals a resonant frequency of \SI{336.1}{\mega\hertz} and a loaded quality factor of \num{65}. 
    (\textbf{c}), Schematic of charge sensing. Loss of an electron, indicated by a change of charge state from $q^0$ to $q^+$, induces a shift of the \absVR-$V_{\mathrm{g}}$ curve from the light blue curve to the dark blue. By choosing a gate voltage on the slope of the curve, $V_{\mathrm{m}}$ , the reflected amplitude at $\lvert V_{\mathrm{R}}(q^0)\rvert$ or $\lvert V_{\mathrm{R}}(q^+)\rvert$ can be used to read out the charge state.}
    \label{Fig1}
\end{figure}
The device used in this work was a Si fin field-effect transistor (FinFET), as illustrated in Fig.~\ref{Fig1}(a). The FinFET consisted of a crystalline Si nanowire channel (width = \SI{35}{\nm}, length = \SI{130}{\nm}, height = \SI{60}{\nm}) with a dielectric coating, which was surrounded by a polycrystalline Si gate on three sides. The device was placed on a cold stage that was operated at \SI{3.9}{\K}, so single quantum dots (QDs) could form in the channel under sub-threshold gate voltages\cite{sellier_subthreshold_2007} and work as charge sensors. A small QD filled with a few electrons can detect loss and gain of a single electron occurring more than \SI{100}{\nm} away\cite{hanson_spins_2007}.

The experimental setup had both fiber-optic access for optically exciting single Er\tplus ions and electrical access for detecting signals from the device, as outlined in Fig.~\ref{Fig1}a. Electrical access via twisted-pair and coaxial cables allowed both DC current and RF reflectometry measurements. The DC and RF measurements shared the same electrical connection to the source of the device. However, they could be performed simultaneously without interference, as the low-frequency DC signal and the high-frequency RF signal were separated by a bias tee consisting of a resistor ($R_1$) and a capacitor ($C_1$). 

The analog bandwidth of the DC measurement was limited to about \SI{2}{\kHz} by the device resistance,  the line capacitance of the twisted-pair cables ($C_{\mathrm{w}}$), and the input impedance of the current pre-amplifier. Therefore, due to the limited bandwidth, the DC measurement was only used for initial device characterization and as a comparison with the RF signal.

The RF reflectometry measurement relied on a resonant circuit formed of an inductor ($L$) connected to the source of the device and the parasitic capacitance ($C_\mathrm{p}$). A carrier RF signal ($V_\mathrm{C}$) was sent to the resonant circuit, and the reflected RF signal ($V_\mathrm{R}$) was measured. A typical circuit characteristic under the optimal impedance matching condition is shown in Fig.~\ref{Fig1}(b), revealing a resonant frequency of $f_\mathrm{r} = 336.1$ \si{\MHz} and a loaded quality factor of $Q_\mathrm{r} = 65$. While the inductance and the parasitic capacitance primarily determine the resonant frequency as $f_\mathrm{r} = 1/(2\pi\sqrt{LC_\mathrm{p}})$, the reflected signal is sensitive to small changes in the device impedance thanks to the high quality factor. The resonant circuit's bandwidth follows $B_\mathrm{r}=f_\mathrm{r}/Q_\mathrm{r}$, and this sets an upper limit of \SI{5.2}{\MHz} on the measurement bandwidth. Eventually, a low-pass filter was used to set a bandwidth limit of \SI{2}{\MHz} on the measured RF signal, ensuring a sufficiently low noise level and a fast response. (See Methods and Supplementary~S1 for details on the measurement setup.)

In order to utilize the RF reflectometry for fast charge sensing, the device was operated under a sub-threshold gate voltage so that the device impedance depends on single electron tunnelling through a QD. This results in a sharp peak in the reflected RF signal as a function of the gate voltage, as illustrated by the colored curves in Fig.~\ref{Fig1}(c). The peak position is highly sensitive to the electrostatic environment in the vicinity of the QD. When the charge state in the vicinity of the QD changes between $q^0$ and $q^+$, the RF signal from the QD switches between the light and the dark blue curves, respectively. In a charge sensing measurement, the gate voltage is typically set on the side of a peak to achieve high sensitivity (e.g. $V_\textrm{m}$ in Fig.~\ref{Fig1}(c)), and the measured \absVR indicates whether the trap is ionized ($q^+$) or neutral ($q^0$).

\begin{figure}
    \centering
    \includegraphics[width=1\textwidth]{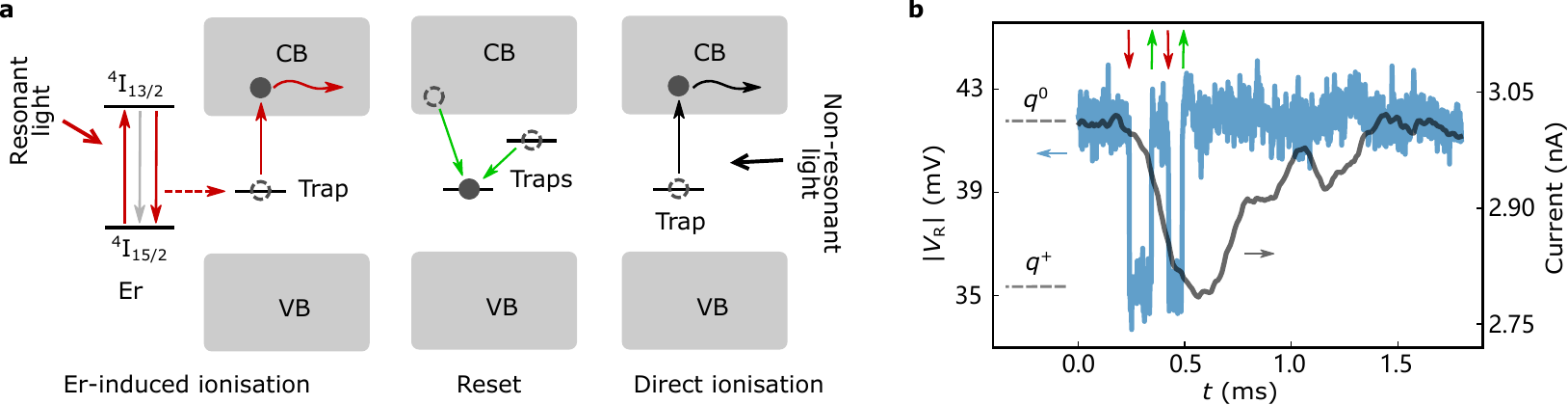}
    \caption{\textbf{Principle of photoionization detection of an Er\tplus ion.} 
    (\textbf{a}), Relevant processes of the photoionization detection of a single Er\tplus ion. The red arrows represent the processes involved in the Er-induced ionization, and the green arrows represent the reset of the ionized trap. First, resonant illumination drives an Er\tplus ion from the \erg ground state to the \ere optically excited state. Then, due to a non-radiative decay process, an electron bound to a trap is promoted into the conduction band (CB) of Si and subsequently escapes (curved red arrow). The grey arrow represents both the radiative decay processes and other non-radiative decay processes that do not cause the trap to ionize. Finally, the ionized trap can capture an electron from the CB or other localized states (green arrows) and reset to its neutral state. Additionally, non-resonant light can directly ionize the trap (black arrows). Here VB is the valence band of Si. 
    (\textbf{b}), An amplitude-time (\absVR-$t$) trace (blue) and a current-time trace (black) recorded simultaneously when the device was under resonant CW illumination. The high level in both traces indicates that the trap is in the neutral state ($q^0$), and the low level indicates that the trap is ionized ($q^+$). The red and green arrows indicate ionization and reset events, respectively.}
    \label{Fig2}
\end{figure}

The photoionization detection involves charge sensing of the ionization of a single trap following resonant excitation and relaxation of an Er\tplus ion\cite{yang_spectral_2022} as illustrated in Fig.~\ref{Fig2}(a). 
The Er\tplus ion can be excited by resonant light from the \erg ground state into the \ere excited state, and the previous study revealed a single-photon excitation process\cite{yang_spectral_2022}. Then it decays back to the ground state via either a radiative or non-radiative process. Some non-radiative decay processes can promote an electron bound to a trap into the conduction band, and the electron subsequently escapes. Afterwards, the trap resets to its initial charge state by capturing an electron from a nearby electron reservoir or other localized states. In addition, non-resonant light can also induce direct ionization of the trap, but this occurs on a much slower time scale than the Er-induced ionization process and does not display a wavelength dependence. The ionization and reset events can be detected via charge sensing as described further below.


Figure.~\ref{Fig2}(b) shows two consecutive pairs of ionization and reset events indicated by the red and green arrows. The RF and DC signals were recorded simultaneously when the device was under resonant continuous-wave (CW) illumination. In the RF signal, an Er-induced ionization event (red arrow) induced a signal drop from $\lvert V_{\mathrm{R}}(q^0)\rvert$ to $\lvert V_{\mathrm{R}}(q^+)\rvert$, and shortly afterwards the signal returned to the original level ($\lvert V_{\mathrm{R}}(q^0)\rvert$) as the trap reset stochastically (green arrow). In comparison, the DC current also shows a response to the four events consistent with the RF signal, but the bandwidth-limited features in the DC signal prohibit accurate identification of these events. In addition, the RF signal can pick up ionization events that reset too rapidly to be recognized from the DC signal. By scanning the laser frequency and detecting the ionization events, two Er\tplus transitions were identified in this device. They can be selectively excited using their specific resonant frequencies. The zero-field frequency is \SI{195054.0}{\GHz} (\SI{1536.972}{\nm}) for Er1, and \SI{195940.7}{\GHz} (\SI{1530.016}{\nm}) for Er2. See Supplementary~S3 for the Zeeman splitting spectra of the two Er\tplus transitions.

\begin{figure}
    \centering
    \includegraphics[width=1.0\textwidth]{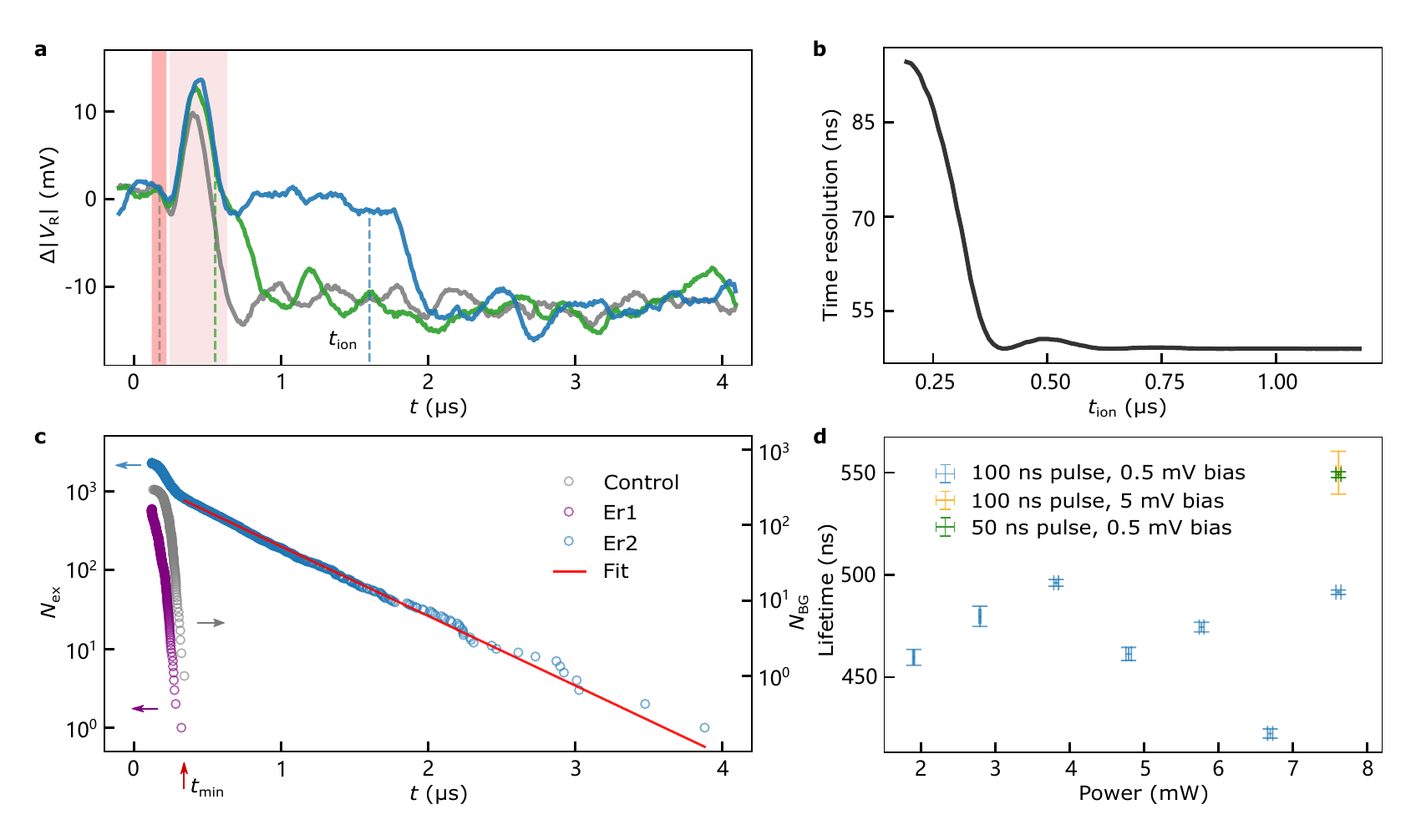}
    \caption{\textbf{Lifetime measurement.}
    (\textbf{a}), Three typical \deltaabsVR-$t$ traces measured from Er2. Here, a relative amplitude \deltaabsVR is calculated by subtracting the averaged value of a \SI{5}{\micro\second} time trace prior to the laser pulse from the measured amplitude \absVR. Dashed lines indicate the fitted ionization times, $t_{\mathrm{ion}}$. The region shaded in pink corresponds to the duration of laser pulse, and the region shaded in light pink highlights a 'laser impact' time window in which a laser-induced transient jump appears.
    (\textbf{b}), Time resolution of the photoionization detection as the function of $t_{\mathrm{ion}}$. The small hump at \SI{0.5}{\micro\second} is caused by the low-pass filtering.
    (\textbf{c}), Statistics of ionization events measured from Er1 (purple), Er2 (blue), and background ionization events (grey). $N_\mathrm{ex}(t)$ and $N_\mathrm{BG}(t)$ represent the number of cycles in which the trap remains neutral at time $t$ among all observed photoionization cycles. 
    An exponential fit (red line) to selected Er2 events ($t_{\mathrm{ion}} > t_{\mathrm{min}}$) gives an excited state lifetime of $\tau_{\mathrm{ex}} =$ \SI{492(1)}{\nano\second} for Er2.
    (\textbf{d}), $\tau_{\mathrm{ex}}$ of Er2 measured at different laser powers (blue), a different pulse length (green), and a different source-drain bias voltage (orange). No significant impact on $\tau_{\mathrm{ex}}$ is observed. The error bars of x-axis and y-axis represent the uncertainty of laser power and the fitting error of $\tau_{\mathrm{ex}}$, respectively.}\label{Fig3}
\end{figure}

The photoionization detection can also be carried out with pulsed laser excitation. Figure~\ref{Fig3}(a) shows three typical traces that contain an ionization event. A \SI{7.6}{\mW}-\SI{100}{\ns} laser pulse (shaded in pink in Fig.~\ref{Fig3}(a)) was applied in each measurement cycle. The appearance of a low-level (\SI{-12}{\mV}) signal indicates a trap ionization event in the measurement cycle, similar to the RF signal response observed in the CW measurement as shown in Fig.~\ref{Fig2}(b). In addition, the laser pulse causes a transient signal jump at $t = \SI{0.4}{\micro\second}$ in the `laser impact' time window (shaded in light pink in Fig.~\ref{Fig3}(a)). This phenomenon occurs for all measurement cycles regardless of the laser frequency. The transient jump becomes more significant for longer laser pulses and/or higher laser powers. To maintain a high detection sensitivity, we kept the maximum laser power and pulse length at \SI{7.6}{\mW} and \SI{100}{\ns}, respectively.

In order to determine the ionization time, $t_{\mathrm{ion}}$, we use a model that includes the laser-induced transient jump, the ionization-induced signal drop, and low-pass filtering. The dashed lines in Fig.~\ref{Fig3}(a) indicate the fitted ionization times for the three traces. When an ionization event occurs after the `laser impact' time window, a well-isolated signal drop can be observed, for example, between \SI{1.8}{\us} and \SI{2.0}{\us} on the blue trace. In contrast, the ionization-induced signal drop on the green curve slightly overlaps with the falling edge of the transient jump as the ionization event occurred shortly after the laser pulse, while the ionization-induced signal drop on the grey curve merges with the transient jump as the ionization event occurred during the laser pulse.

We then analyze the time resolution of the photoionization detection. The noise in the RF signal and the fluctuation of the ionization-induced signal drop directly affect the stability of the ionization-induced falling edge and, thus, the ionization time. These two factors contribute to \SI{56}{\ns} time resolution. Additionally, the laser-induced transient jump can influence the fitting accuracy of the ionization time for the events occurring during or shortly after the laser pulse. By considering all the above effects, the time resolution of the photoionization detection is analyzed and plotted as a function of ionization time, $t_{\mathrm{ion}}$, in Fig.~\ref{Fig3}(b). Details of fitting and simulation can be found in Supplementary~S6.


To investigate the excited state lifetime, pulsed detection was performed for Er1 and Er2 with \SI{7.6}{\mW}-\SI{100}{\ns} resonant laser pulses. The short excitation pulse resulted in a relatively low ionization probability of $\sim$\num{6e-4} and $\sim$\num{3e-4} event per measurement cycle for Er1 and Er2 and \num{594} and \num{2280} ionization events were recorded, respectively. These events include Er-induced ionization events and frequency-independent background events. A control experiment was carried out with \SI{7.6}{\mW}-\SI{100}{\ns} non-resonant laser pulses, and \num{294} background events were recorded at an ionization probability of $\sim$\num{6e-5} event per cycle.

In the pulsed detection, the Er\tplus ion can be excited into the \ere excited state by a laser pulse and then relax back to the \erg ground state with a certain probability of inducing an ionization event. The statistics of the Er-induced ionization events should follow an exponential decay function: $N_{\mathrm{ex}}(t) = N_{\mathrm{ex}}(t_0) \times e^{(t_0-t)/\tau_{\mathrm{ex}}}$, where $N_{\mathrm{ex}}(t)$ is the number of cycles in which the Er\tplus ion is in the excited state at time $t$, and $\tau_{\mathrm{ex}}$ is the excited state lifetime.

Statistics of the ionization events from the three experiments are shown in Fig.~\ref{Fig3}(c). All the ionization events from the control experiment (grey) occurred during or shortly after the laser pulse. Therefore, the latest ionization time in the control experiment can be used as a minimum threshold ($t_{\mathrm{min}} = $ \SI{343}{\ns}) to exclude the background events. As shown in Fig.~\ref{Fig3}(c), \num{819} ionization events occurring after $t_{\mathrm{min}}$ are selected from the Er2 measurement (blue), and an exponential decay fit (red line) to these events gives an excited state lifetime of Er2, $\tau_{\mathrm{ex}}=$ \SI{492(1)}{\ns}. This threshold, $t_{\mathrm{min}}$, is also used to exclude background events for generating the Zeeman splitting spectrum of Er2 (Fig.~S3b). In contrast, all the ionization events from the Er1 measurement (purple points in Fig.~\ref{Fig3}(c)) occurred before $t_{\mathrm{min}}$. As a result, the background events cannot be excluded using the selection method. Nevertheless, we can estimate the background count in the Er1 measurement based on the ionization probability in the control experiment. Approximately, \SI{90}{\percent} of the \num{594} events are expected to be Er-induced ionization events. We infer that the excited state lifetime of Er1 is shorter than \SI{50}{\ns} but cannot determine it accurately due to the limited time resolution of the detection for these events.

We then investigated the effects of laser power, pulse length and source-drain bias voltage on the excited state lifetime of Er2. These factors may influence the local environment in the channel of the device, such as the amount of photo-induced carriers and the electric field, which may affect the excited state lifetime of an Er\tplus ion. We first varied the laser power from \SI{1.9}{\mW} to \SI{7.6}{\mW} at a pulse length of \SI{100}{\ns} and a bias voltage of $V_{\mathrm{s}} =$ \SI{0.5}{\mV}, and the results are plotted as blue markers in Fig.~\ref{Fig3}(d). While keeping the laser power fixed at \SI{7.6}{\mW}, we shortened the pulse length from \SI{100}{\ns} to \SI{50}{\ns} in one experiment (green marker) with $V_{\mathrm{s}} =$ \SI{0.5}{\mV} and increased the bias voltage to $V_{\mathrm{s}} =$ \SI{5.0}{\mV} in another experiment (orange marker) with a pulse length of \SI{100}{\ns}. Overall, the laser power, the pulse length, or the source-drain bias voltage does not affect the excited state lifetime of Er2 significantly, as shown in Fig.~\ref{Fig3}(d). See Supplementary~S8 for detailed experimental and fitting results.

\begin{figure}
    \centering
    \includegraphics[width=0.5\textwidth]{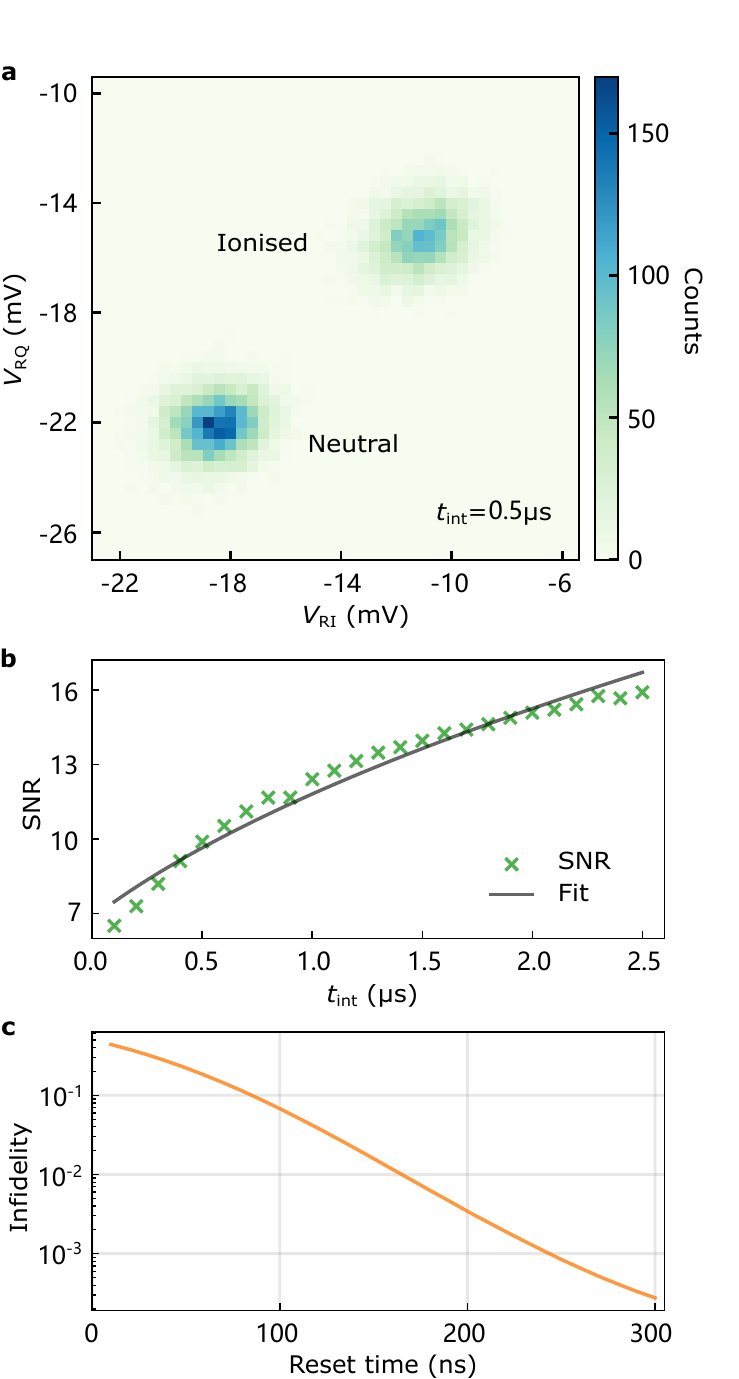}
    \caption{\textbf{Signal-to-noise ratio (SNR) and detection fidelity.}
    (\textbf{a}), A 2D histogram of the reflected RF signal $V_\mathrm{R}$ in the $IQ$ plane with an integration time of \SI{0.5}{\us}. $V_\mathrm{RI}$ and $V_\mathrm{RQ}$ stand for the in-phase and quadrature components of $V_\mathrm{R}$, respectively. Two well separated regions correspond to the ionized and neutral states of the trap.
    (\textbf{b}), SNR as a function of integration time. The black curve is a fit with $\mathrm{SNR}=\mathrm{SNR}_{\SI{1}{\micro\second}} \times \sqrt{(t_0+t_{\mathrm{int}})/ \SI{1}{\micro\second}}$, where $\mathrm{SNR}_{\SI{1}{\micro\second}}$ is a characteristic SNR with an effective integration time of \SI{1}{\us} and $t_0$ is an intrinsic integration time.
    (\textbf{c}), Infidelity of the photoionization detection as a function of reset time.}
    \label{Fig4}
\end{figure}

In the final section, we analyzed the sensitivity and limits of the RF detection, including the signal-to-noise ratio (SNR) of the RF reflectometry and fidelity of the photoionization detection, which are vital to enable the fast detection of optical centers.

In the present study, the reflected RF signal, $V_\mathrm{R}$, is mainly sensitive to changes in the device resistance, which is typical when the resonant circuit is connected to the source or drain. In general, a change of charge state can induce both resistive and reactive responses, which lead to changes in the amplitude and phase of $V_\mathrm{R}$. To fully characterize the SNR, the in-phase and quadrature components of $V_\mathrm{R}$ are plotted into a 2D histogram in the $IQ$ plane. Figure~\ref{Fig4}(a) shows the reflected RF signals before and after an ionization event with a signal integration time of $t_{\mathrm{int}}=$ \SI{0.5}{\us}. Two well-separated regions correspond to the ionized and neutral states of the trap. The SNR is calculated as $\mathrm{SNR} = \lvert \Delta V_\mathrm{R}\rvert/\sigma_V$, where $\lvert \Delta V_\mathrm{R}\rvert$ is the center-to-center distance between the two regions, and $\sigma_V$ is their averaged standard deviation. The SNR values calculated with different integration times are shown in Fig.~\ref{Fig4}(b). This dependence can be described by the following equation, $\mathrm{SNR}=\mathrm{SNR}_{\SI{1}{\micro\second}} \times \sqrt{(t_0+t_{\mathrm{int}})/ \SI{1}{\micro\second}}$\cite{barthel_fast_2010}, and the fit (black line in Fig.~\ref{Fig4}(b)) yields $\mathrm{SNR}_{\SI{1}{\micro\second}} = 9.6$, a characteristic SNR for an effective integration time of \SI{1}{\us}, and $t_0 = \SI{0.50(6)}{\micro\second}$, an intrinsic integration time corresponding to the measurement bandwidth of \SI{2}{\MHz}.

The fidelity of the photoionization detection is influenced by the SNR of the RF reflectometry and the duration of the ionization signal, equivalent to the reset time.
In the present device, the SNR with any integration time above \SI{0.5}{\us} exceeds 10, and the reset time constant is \SI{70.9+-0.1}{\us}. As a result, most ionization events can be detected with near-unity fidelity. Nevertheless, \SI{0.7}{\percent} of all ionization events are expected to reset within \SI{0.5}{\us} at the limit of the detection bandwidth. These short-lived events would have attenuated ionization signal contrast and consequently could be missed in the detection.
In order to quantify this impact, we simulated the infidelity of photoionization detection for short-lived ionization events, as shown in Fig.~\ref{Fig4}(c). For each reset time, the infidelity is calculated using the expected signal amplitude and the measured noise. (See Supplementary~S7 for details.)
The simulation results reveal \SI{99}{\percent} detection fidelity for ionization events with a duration longer than \SI{170}{\ns}, and overall detection fidelity of \SI{99.96}{\percent} for all ionization events in the present device.

\section*{Discussion}
We have demonstrated fast photoionization detection of single Er\tplus ions in a Si nano-transistor using RF reflectometry. At an analog bandwidth of \SI{2}{\MHz}, this technique provides a SNR of \num{9.6} with an integration time of \SI{0.5}{\us} and photoionization detection time resolution of below \SI{100}{\ns}. The fast detection enables \ere excited state lifetime measurements for single Er\tplus ions in a Si nano-transistor.

The SNR and time resolution of RF readout can be further improved. Firstly, a lower-noise first-stage amplifier could further reduce the noise level of the measured RF signal, for example, a superconducting quantum interference device (SQUID) amplifier\cite{JPA1} or a Josephson parametric amplifier\cite{stehlik_fast_2015,schaal_fast_2020}. Secondly, replacing the surface mount inductor with a superconducting inductor\cite{SC_L} can minimize dissipative losses and reduce the parasitic capacitance, enhancing the signal contrast. However, using a superconducting inductor constrains high magnetic fields to be in the plane of the superconducting film, which should be considered in device design. As an example, robust charge sensitivity has been demonstrated in moderate in-plane magnetic fields up to \SI{1}{T}\cite{SC_L}. Finally, part of the signal fluctuations on a time scale of microseconds are likely due to light-induced heating and random photocarrier generation in the leads of the FinFET device, as a large area of the leads around the channel was illuminated by the divergent light in the present study. This impact can be reduced in new device structures with an Er-doped region separated from the channel. Also, undesirable illumination and laser heating can be further reduced by coupling the Er-doped region to an optical cavity or waveguide\cite{xu_ultrashallow_2021}.

RF reflectometry provides a promising direction forward for scalable optical quantum systems. The RF resonant circuit was connected to the source of the device in the present study; as a result, a dissipative response, i.e. the reflected amplitude, was used for detection. Alternatively, the resonant circuit can be connected to the gate for dispersive readout, which has been implemented on single lead QD devices\cite{house_high_sensitivity_2016}. This structural simplification, combined with channel multiplexing techniques\cite{hornibrook_frequency_2014}, provides an attractive solution to the readout and control of single optical centers in large-scale integrated systems.

\section*{Methods}
The device was mounted on a printed circuit board (PCB) and placed on a cold stage that was operated at \SI{3.9}{\K} in a liquid helium free cryogenic system. For the RF measurement, a carrier signal from an RF signal generator was split into two parts, one sent to the device and the other used for signal demodulation. The first part passed through room temperature (RT) attenuators (\SI{-36}{\dB}), cryogenic attenuators (\SI{-40}{\dB}), and a directional coupler and then reached the device PCB. The reflected RF signal went through the directional coupler via a different path from the incoming RF signal and then was amplified by a commercial cryogenic amplifier, a homemade cryogenic amplifier and an RT amplifier sequentially. The amplified signal and the second part of the original RF signal were sent to a quadrature demodulator for homodyne demodulation. Finally, the demodulated signals, $V_{\mathrm{RI}}$ and $V_{\mathrm{RQ}}$, were sent to an oscilloscope after passing a \SI{2}{\MHz} low-pass filter. 

The laser pulses were generated by an acousto-optic modulator that was controlled by a pulse generator. A small portion of the pulsed light was sent to a photodetector, and the output voltage was sent to the oscilloscope for triggering and power monitoring. Specifically, the rising edge of the laser pulse captured by the photodetector was used as the trigger in the present study and corresponds to $t=0$. Due to instrumental response times and latencies of optical fibers and coaxial cables, the arrival time of the laser pulse at the device is estimated to be $t = 30 \si{\nano\second}$. The induced RF signal at this specific moment should start to appear in the oscilloscope signal at $t=75\si{\nano\second}$ but actually shows up much later (about $t=0.4\si{\micro\second}$ in Fig.\ref{Fig3}(a)) due to the \SI{2}{\MHz} low-pass filtering. These time delays remained stable within \SI{+-1}{\ns} and should not affect the analysis of lifetime and time resolution, as we did not change the lengths of fibers and cables throughout the present study. More details about the connections can be found in Supplementary~S1\nocite{Varactor,time_jitter}. 

\section*{Supplementary materials}
Supplementary material for this article is available at (link).

\bibliography{scibib}

\bibliographystyle{Science}


\textbf{Acknowledgments:} We acknowledge the AFAiiR node of the NCRIS Heavy Ion Capability for access to ion-implantation facilities. 
\textbf{Funding:} This work was supported by the National Key R\&D Program of China (Grant No. 2018YFA0306600), Anhui Initiative in Quantum Information Technologies (Grant No. AHY050000), and Anhui Provincial Natural Science Foundation (Grant No. 2108085MA15). 
\textbf{Author contributions:} C.Y. conceived the project and guided the experimental investigation and analysis. Y.Z. carried out the experiments and performed the numerical analysis. Y.Z., W.F., and J.Y. contributed to building the experimental setup. B.C.J. and J.C.M. designed and performed the implantation. Y.Z. and C.Y. wrote the manuscript with input from other authors.
\textbf{Competing interests:} Authors declare that they have no competing interests.
\textbf{Data and materials availability:} Source data for the plots are available at (link provided before publication). All data needed to evaluate the conclusions in the paper are present in the paper and/or the Supplementary Materials.

\end{document}